\documentclass[amsmath,amssymb,11pt,pre]{revtex4-1}

\usepackage{graphicx}
\usepackage{dcolumn}
\usepackage{bm}
\usepackage{color}%
\usepackage{subcaption}

\usepackage[normalem]{ulem}

\begin{document}


\title{Surface gravity waves propagating in a rotating frame: the Ekman-Stokes instability}

\author{Kannabiran Seshasayanan}
\email{kannabiran.seshasayanan@cea.fr}
\author{Basile Gallet}%
 \affiliation{%
 Service de Physique de l'{\'E}tat Condens{\'e}, CEA, CNRS UMR 3680, Universit{\'e} Paris-Saclay,
CEA Saclay, 91191 Gif-sur-Yvette, France
 }%


\begin{abstract}

We report on an instability arising when surface gravity waves propagate in a rotating frame. The Stokes drift associated to the uniform wave field, together with global rotation, drives a mean flow in the form of a horizontally invariant Ekman-Stokes spiral. We show that the latter can be subject to an instability that triggers the appearance of an additional horizontally-structured cellular flow. We determine the instability threshold numerically, in terms of the Rossby number $\text{Ro}$ associated to the Stokes drift of the waves and the Ekman number $\text{E}$. We confirm the numerical results through asymptotic expansions at both large and low Ekman number. At large $\text{E}$ the instability reduces to that of a standard Ekman spiral driven by the wave-induced surface stress instead of a wind stress, while at low $\text{E}$ the Stokes-drift profile crucially determines the shape of the unstable mode. In both limits the instability threshold asymptotes to an Ekman-number-independent critical Rossby number, which in both cases also corresponds to a critical Reynolds number associated to the Lagrangian base-flow velocity profile. Parameter values typical of ocean swell fall into the low-$\text{E}$ unstable regime: the corresponding ``anti-Stokes'' flows are unstable, with possible consequences for particle dispersion and mixing.

\end{abstract}

\pacs{Valid PACS appear here}
\maketitle

%
The interaction between waves and mean flows is a key ingredient of oceanic and atmospheric dynamics. Focusing on the ocean surface only, mean flows can refract and focus surface gravity waves \cite{Phillips,BuhlerBook,Dysthe2001,Kenyon,Gallet2014,Gutierrez}, while surface gravity waves can induce and distort near-surface currents \cite{CraikBook,Teixeira2002,Phillips94,Phillips05,Humbert}. A particularly visual example of wave-induced currents are Langmuir cells, i.e., cellular motion near the ocean surface, the axis of the cells being aligned with the dominant direction of wave propagation. These wave-induced mean flows affect  dispersion and mixing in the upper ocean \cite{McWilliams97,McWilliams1999}. With the goal of describing the emergence of Langmuir cells, Craik and Leibovich \cite{Craik1976,Leibovich80,Craik1982} derived an equation governing the evolution of a slow incompressible mean flow ${\bf u}$ arising at second order in the weak slope of the surface gravity waves:
\begin{align}
\partial_t { \bf u} = - \nabla p + ( {\bf u} + {\bf u}_s ) \times {\bm \omega} \, . \label{eq:CL_standard}
\end{align}
In this equation, ${\bf u}_s$ denotes the Stokes drift associated with the wave field, ${\bm \omega}={\bm \nabla} \times {\bf u}$ is the vorticity field and $p$ denotes the generalized pressure field. This equation resembles the standard Navier-Stokes equation, with the addition of a ``vortex force'' term ${{\bf u}_s\times {\bm \omega}}$. Focusing on the interaction between a spanwise-varying Stokes drift and a vertically sheared wind-driven current, Craik and Leibovich showed that the vortex force induces cellular motion reminiscent of Langmuir cells. Craik later realized that, even when the Stokes drift is horizontally homogeneous, an instability mechanism leads to the spontaneous emergence of a similar cellular flow \cite{Craik1977}. A complication arising in Craik's study -- as well as in subsequent ones -- is that the wind stress imposed at the fluid surface induces unbounded growth of the kinetic energy of the background shear flow, as momentum accumulates inside the semi-infinite fluid region. These studies therefore need to restrict attention to a ``quasi-stationary'' base state, where the slowly evolving mean flow is approximated by a steady one and standard linear stability analysis can be applied. This approach is sound when the instability unfolds on a timescale much faster than the evolution of the base flow \cite{Craik1977, tsai2015, tsai2017}, an assumption that breaks down in the vicinity of the instability threshold.

As we will see in the following, a solution to this issue consists in including global rotation -- a key ingredient of oceanic flows -- to arrest the accumulation of momentum in the fluid layer. Wave mean-flow interactions are modified significantly when global rotation ${\bm \Omega}$ and viscosity $\nu$ are taken into account, both at second-order in wave slope \cite{Huang}. First, the wave-averaged equation for the slowly evolving mean flow becomes:
\begin{align}
\partial_t { \bf u} = -  \nabla p + ( {\bf u} + {\bf u}_s ) \times ( {\bm \omega} + 2 {\bm \Omega}) + \nu \Delta {\bf u} \, .\label{eq:CL_rotatingframe}
\end{align}
Secondly, viscosity crucially modifies the boundary conditions for the mean flow at the free surface \cite{LH53,Unluata,Madsen,Xu}. On average over a wave period, the stress-free boundary condition at the deformed top boundary results in a net viscous stress $\rho \nu \partial_z {{ \bf u}_s}|_{z=0}$ acting on the underlying mean flow. Balancing this wave-induced stress with the standard viscous stress $\rho \nu \partial_z {\bf u}_{\perp}|_{z=0}$ exerted by the fluid slightly below the surface yields the modified boundary conditions for the mean flow at the surface:
\begin{align}
\partial_z  { \bf u}_{\perp}|_{z=0}  = \partial_z {{ \bf u}_s}|_{z=0} \, ; \qquad {\bf e}_z \cdot {\bf u}|_{z=0} = 0 \, , \label{eq:stress}
\end{align}
where $z$ is the vertical direction, with $z=0$ the rest position of the free surface, ${\bf e}_z$ is the unit vector along $z$ and ${ \bf u}_{\perp}$ is the horizontal velocity vector. 
We stress the fact that including a small viscosity $\nu$ is a very singular perturbation to the undamped problem: the boundary condition (\ref{eq:stress}) is independent of $\nu$ and does not reduce to the standard stress-free boundary condition $\partial_z {\bf u}_{\perp}|_{z=0} = 0$ when $\nu \to 0$. As far as we can tell, this boundary condition was initially derived more than half a century ago by Longuet-Higgins \cite{LH53}. Surprisingly however, even though some studies correctly include it \cite{Unluata,Madsen,Xu}, many studies in oceanography and physics forget about the wave-induced stress. Part of the reason may be that the original derivation by Longuet-Higgins is rather involved mathematically: it requires switching to a curvilinear frame attached to the wavy interface before performing a boundary-layer approximation. With the goal of popularizing the modified boundary condition, Xu \& Bowen provide a much simpler derivation, that they refer to as an energy budget \cite{Xu}. However, their approach really is a momentum budget: we provide a similar derivation in appendix  \ref{appBC}, correcting a few incorrect steps in Xu \& Bowen.

Most other studies motivated by the oceanographic situation forget about the wave-induced stress near the surface. This can be justified when a wind stress is included in the problem, provided it is much stronger than the wave-induced stress. If the latter is evaluated using the molecular viscosity of water, then one concludes that any realistic wind stress dominates the wave-induced surface stress by orders of magnitude. However, instead of molecular viscosity, many of these studies consider eddy viscosity all the way to the fluid surface. For instance, the seminal papers of Craik \& Leibovich \cite{Craik1976} and Huang \cite{Huang} assume values of the eddy viscosity of the order of $10^{-2}$m$^2$s$^{-1}$, but forget about the wave-induced stress. Consider typical parameter values for swell waves: a surface Stokes drift $U_s=0.068$m.s$^{-1}$, a wavenumber $k=0.105$m$^{-1}$ and a typical wind stress of $0.037$N.m$^{-2}$ \cite{McWilliams97}. From these values, one concludes that the wave-induced stress estimated using eddy viscosity is four times larger than the wind stress, and thus cannot be neglected. Whether molecular or eddy viscosity should be retained in the vicinity of the free surface is debatable and we shall not elaborate more here. Instead, in the following we consider $\nu$ to be the (uniform) viscosity of the fluid and focus on propagating waves in the absence of external wind stress. 

An important consequence of both viscosity and global rotation is that ${\bf u}= {\bm 0}$ is a solution to equation (\ref{eq:CL_standard}) but not to equation (\ref{eq:CL_rotatingframe}). Indeed, the simplest steady solution to equation (\ref{eq:CL_rotatingframe}) is a horizontally invariant mean flow, driven both by the wave-induced surface-stress (\ref{eq:stress}) and by the body-force term ${\bf u}_s \times 2 {\bm \Omega}$ in equation (\ref{eq:CL_rotatingframe}). This ``Coriolis-Stokes'' body force corresponds to the shearing of planetary vorticity by the Stokes drift \cite{suzukifox,hiuimpac}. The resulting base flow was coined the Ekman-Stokes spiral \cite{polton}: it is somewhat similar to an Ekman spiral, but it is driven by the Stokes drift and the wave-induced surface stress, instead of a surface wind stress. 
The stability of the Ekman-Stokes spiral has received little attention so far, except for one study \cite{Gnanadesikan}. The authors of that study include a wind-stress, but they omit the wave-induced surface stress and use a Galerkin expansion that is incompatible with the boundary conditions of the problem, making any quantitative comparison difficult. 

In the following, we restrict attention to surface waves propagating in a rotating frame in the absence of wind stress, with the following questions in mind: is the Ekman-Stokes spiral stable? Can it break down into cellular motion? How much mixing and dispersion would it induce then? How much energy can the instability extract from the wave field?

To address these questions, we introduce in section \ref{sec:base_flow} an idealized pilot problem where monochromatic surface waves propagate in a rotating frame and derive the corresponding Ekman-Stokes spiral. In section \ref{sec:linstab}, we consider the linear stability of this base flow. We solve the corresponding eigenvalue problem numerically and demonstrate the existence of an instability. The numerical problem becomes particularly stiff at low Ekman number, which calls for an analytical confirmation of the numerical results. In section \ref{sec:asympt}, we thus consider the asymptotic limits of large and low Ekman number. After connecting the large-Ekman-number regime to the instability of the standard Ekman spiral, we perform a boundary-layer analysis to shed light on the low-Ekman-number instability mechanism, where the Stokes drift profile plays a key role in setting the instability threshold and the structure of unstable eigenmode. In the discussion section \ref{sec:discussion}, we show that the instability threshold -- a critical Rossby number based on the characteristics of the Stokes drift profile -- also corresponds to a critical Reynolds number associated to the Lagrangian velocity profile. We finally discuss the energy budget in the nonlinear regime of the instability, by deriving an upper bound on the power dissipated by the background flow. 

\section{\label{sec:base_flow} Ekman-Stokes spiral}

\begin{figure}
\begin{center}
a.\includegraphics[width=7.0cm]{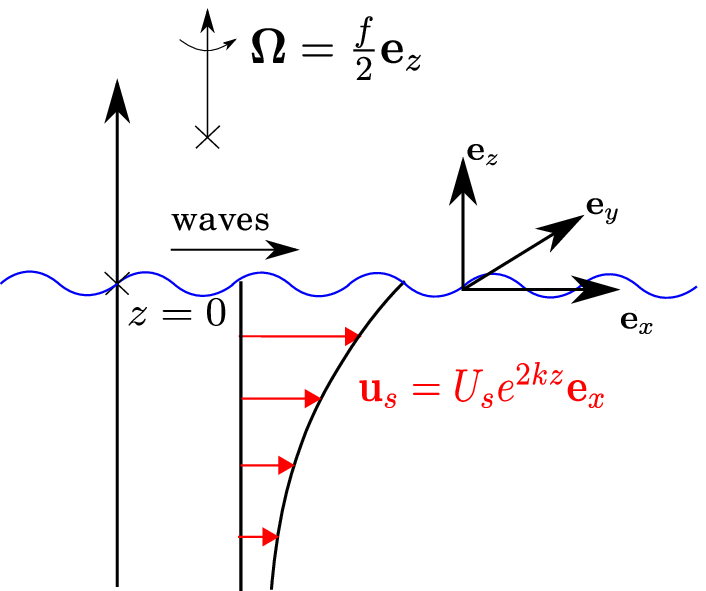}
b.\includegraphics[width=8cm]{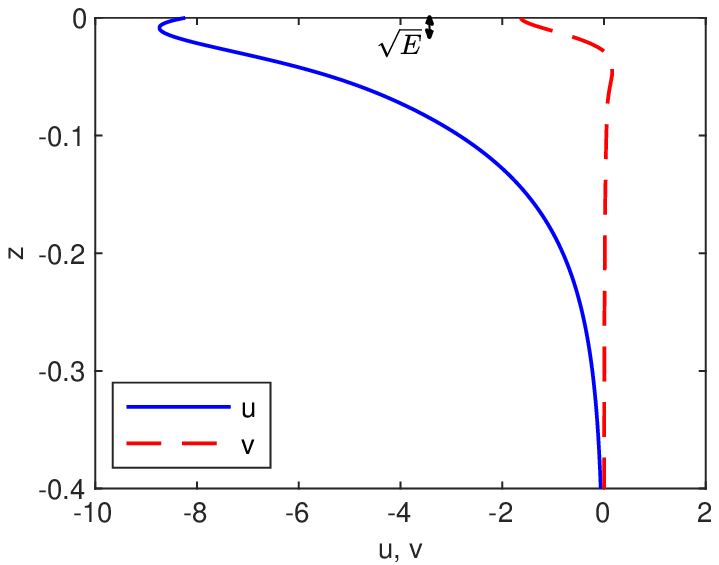}
\end{center}
\caption{\textbf{Left:} Theoretical setup. Monochromatic surface gravity waves propagate towards $x>0$ in a frame rotating at a rate $\Omega=f/2$ around the vertical $z$ axis. \textbf{Right:} The base-flow solution associated with this geometry for $Ro=10$ and $\text{E}=10^{-3}$, in dimensionless units (see text for details). Notice the two characteristic scales in the vertical direction, $1/4\pi$ from the Stokes drift and $\sqrt{\text{E}}$ from the standard Ekman layer scaling.}
\label{fig:schematic}
\end{figure}

The theoretical setup is sketched in Fig. \ref{fig:schematic}a: we consider a layer of incompressible fluid with infinite extent in both horizontal directions $x$ and $y$ and semi-infinite extent along $z$, in a frame rotating at angular frequency $\Omega$ around the vertical axis ${\bf e}_z$. At the free surface is a field of monochromatic surface gravity waves propagating towards positive $x$, with an amplitude $a_w$, a frequency $\sigma$ and a wavenumber $k = 2 \pi/\lambda$, $\lambda$ being the wavelength.    The waves have a weak-slope, $a_w k \ll 1$. We focus on global rotation rates $\Omega$ much lower than the wave frequency $\sigma$, so that we can safely neglect the modifications to the wave dispersion relation induced by global rotation \cite{Backus}. The surface gravity waves thus follow the standard dispersion relation $\sigma=\sqrt{gk}$. The interaction between these waves and a slowly evolving background flow ${\bf u} = u {\bf e}_x + v {\bf e}_y  + w {\bf e}_z $ is governed by equation (\ref{eq:CL_rotatingframe}) above, where the Stokes drift associated to the deep-water monochromatic surface-wave field is:
\begin{align}
{\bf u}_s = u_s(z) {\bf e}_x = U_s e^{2 k z} {\bf e}_x \, , \label{eq:Stokesdrift}
\end{align}
with $U_s = a_{_w}^2 \sigma k$.
The goal of the present study is to investigate the solutions to equation (\ref{eq:CL_rotatingframe}) with the Stokes drift (\ref{eq:Stokesdrift}) and the boundary conditions (\ref{eq:stress}), and the stability of these solutions. The problem involves two dimensionless parameters:
\begin{align}
\text{E} = \nu/(f \lambda^2) \, , \qquad \text{Ro}= U_s/(\lambda f) \, .
\end{align}
$\text{E}$ is a standard Ekman number written in terms of the Coriolis parameter $f=2\Omega$, and $\text{Ro}$ is a Rossby number associated with the Stokes drift velocity $U_s$. Both parameters are based on the wavelength $\lambda$, which is the typical depth of the Stokes drift profile. 

We non-dimensionalize equation (\ref{eq:CL_rotatingframe}) using the lengthscale $\lambda$ and the timescale $f^{-1}$. Denoting the dimensionless quantities with a $*$, we obtain:

\begin{align}
\partial_{t^{*}} { \bf u}^{*} = -  {\bm \nabla}^{*} p^{*} + ( {\bf u}^{*} + {\bf u}_s^{*} ) \times ( {\bm \nabla}^{*} \times {\bf u}^{*} + {\bf e}_z) + \text{E} \, \Delta^{*} {\bf u}^{*} \, , \label{eq:rotatingCLadim}
\end{align}
where ${\bf u}_s^* = \text{Ro}\, e^{4 \pi z^*} {\bf e}_x$. The dimensionless boundary conditions are:
\begin{align}
\partial_{z^*}  { \bf u_{\perp}^*}|_{z^*=0} = 4\pi \, \text{Ro}\, {\bf e_x} \, , \qquad { w^*}|_{z^*=0} = 0 \, . \label{eq:stressadim}
\end{align}
To alleviate the algebra, we drop the stars in the following, all the quantities being dimensionless unless stated otherwise.

A consequence of global rotation is that equation (\ref{eq:rotatingCLadim}) admits a time-independent and horizontally invariant base-flow solution. The incompressibility condition together with the boundary conditions yield $w=0$, while the horizontal velocity components are:
\begin{align}
u = & 4 \pi \sqrt{\text{E}} \sqrt{ \frac{(4 \pi)^4 \text{E}^2 + 4}{(4 \pi)^4 \text{E}^2 + 1} } \cos \left( z/\sqrt{2 \text{E}} - \theta \right) e^{ z/\sqrt{2 \text{E}}} \text{Ro}- \frac{1}{1 + (4 \pi)^4 \text{E}^2} \text{Ro}\, e^{4 \pi z}, \label{eq:base_flow_u}\\
v = & 4 \pi \sqrt{\text{E}} \sqrt{ \frac{(4 \pi)^4 \text{E}^2 + 4}{(4 \pi)^4 \text{E}^2 + 1} } \sin \left( z/\sqrt{2 \text{E}} - \theta \right) e^{ z/\sqrt{2 \text{E}}} \text{Ro} + \frac{(4 \pi)^2 \text{E}}{1 + (4 \pi)^4 \text{E}^2} \text{Ro} \, e^{4 \pi z}. \label{eq:base_flow_v}
\end{align}
where $\theta = \text{arctan}[ ((4 \pi)^4 \text{E}^2 + (4 \pi)^2 \text{E} + 2)/((4 \pi)^4 \text{E}^2  - (4 \pi)^2 \text{E} + 2) ]$. 
This base flow was derived in Lagrangian form by Madsen \cite{Madsen} and is shown in figure \ref{fig:schematic}b. It is referred to as the Ekman-Stokes spiral, as it resembles an Ekman spiral but is driven by the Stokes drift instead of a wind stress. The Stokes drift appears directly in equation (\ref{eq:rotatingCLadim}) but also in the wave-induced surface stress (\ref{eq:stressadim}). As can be seen in figure \ref{fig:schematic}b, the velocity profile evolves over two different length scales in the vertical direction: the Ekman layer depth $\lambda \sqrt{\text{E}} = \sqrt{\nu/f}$, and the wavelength $\lambda$. The former is the length scale of the classical Ekman spiral, while the latter is the signature of the depth dependence of the Stokes drift. In the limit of very small Ekman number, the velocity profile tends to ${\bf u} = - {\bf u}_s$ pointwise. This Eulerian profile is often referred to as an ``anti-Stokes'' flow \cite{Ursell,Pollard,Hasselmann}, as it cancels the Stokes drift velocity in the expression of the Lagrangian velocity, ${\bf u}_L={\bf u}+{\bf u}_s \simeq {\bm 0}$. However, the vertical derivatives of ${\bf u}$ and $- {\bf u}_s$ differ strongly in the vicinity of the fluid surface, because these two fields satisfy different boundary conditions. As a result, there is an Ekman boundary layer extending over a typical depth $\sqrt{\nu/f}$ near $z=0$ for ${\bf u}$ to satisfy the boundary conditions (\ref{eq:stressadim}) at the fluid surface.

\section{\label{sec:linstab} Linear stability analysis}

\begin{figure}
\begin{center}
a.\\
\includegraphics[width=10cm]{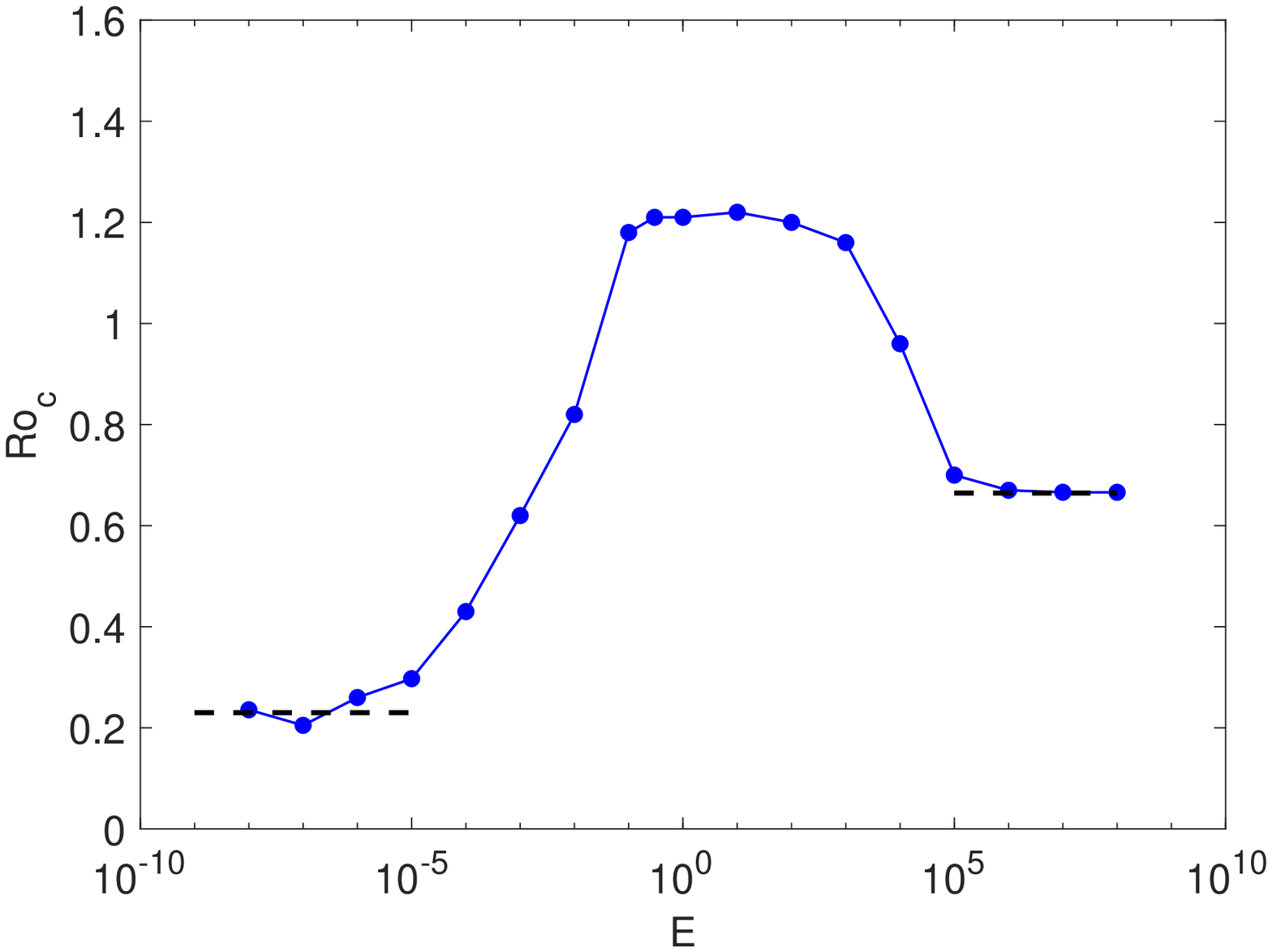} \\
b.\\
\includegraphics[width=10cm]{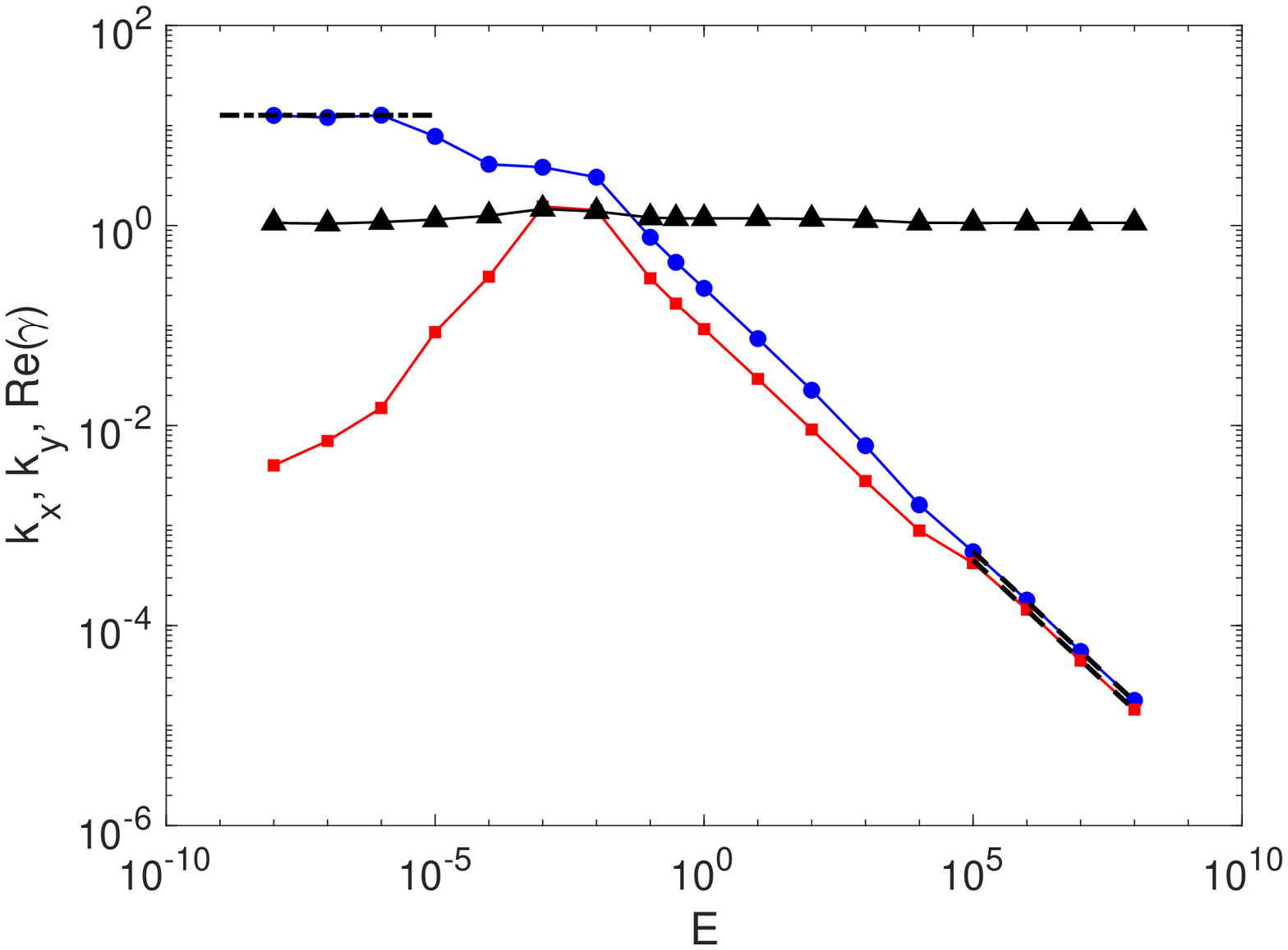}
\end{center}
\caption{a. Critical Rossby number $\text{Ro}_c$ as a function of the Ekman number. b. The wavenumbers $k_x$ ($\bullet$) and $k_y$ ($\square$), and the angular frequency $\text{Re}(\gamma)$ ($\triangle$) at criticality. The dashed lines indicate the asymptotic results for low and large $\text{E}$, which fully confirm the numerical study.}
\label{fig:linstab}
\end{figure}

\begin{figure}
\begin{center}
\includegraphics[width=8cm]{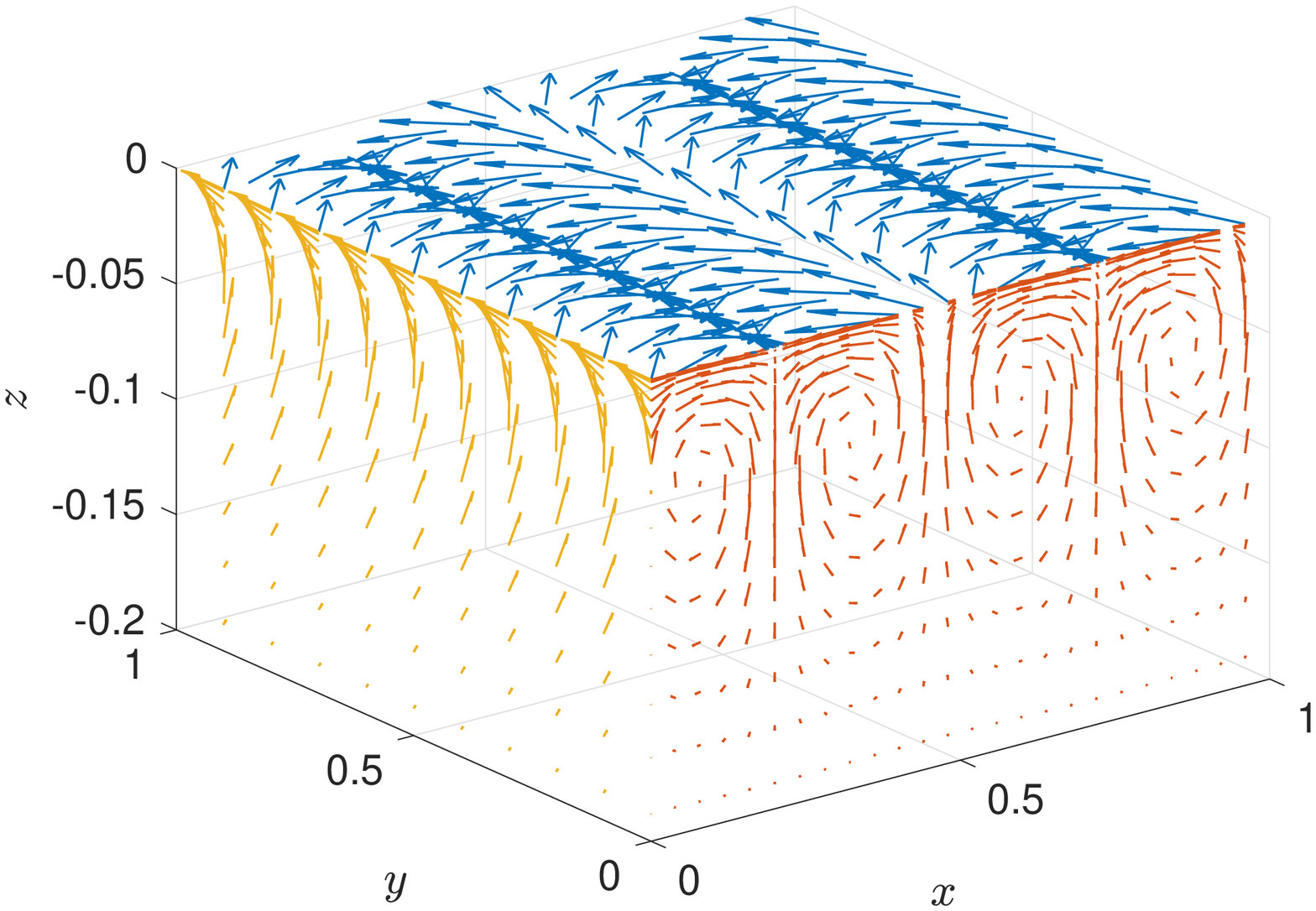}
\includegraphics[width=8cm]{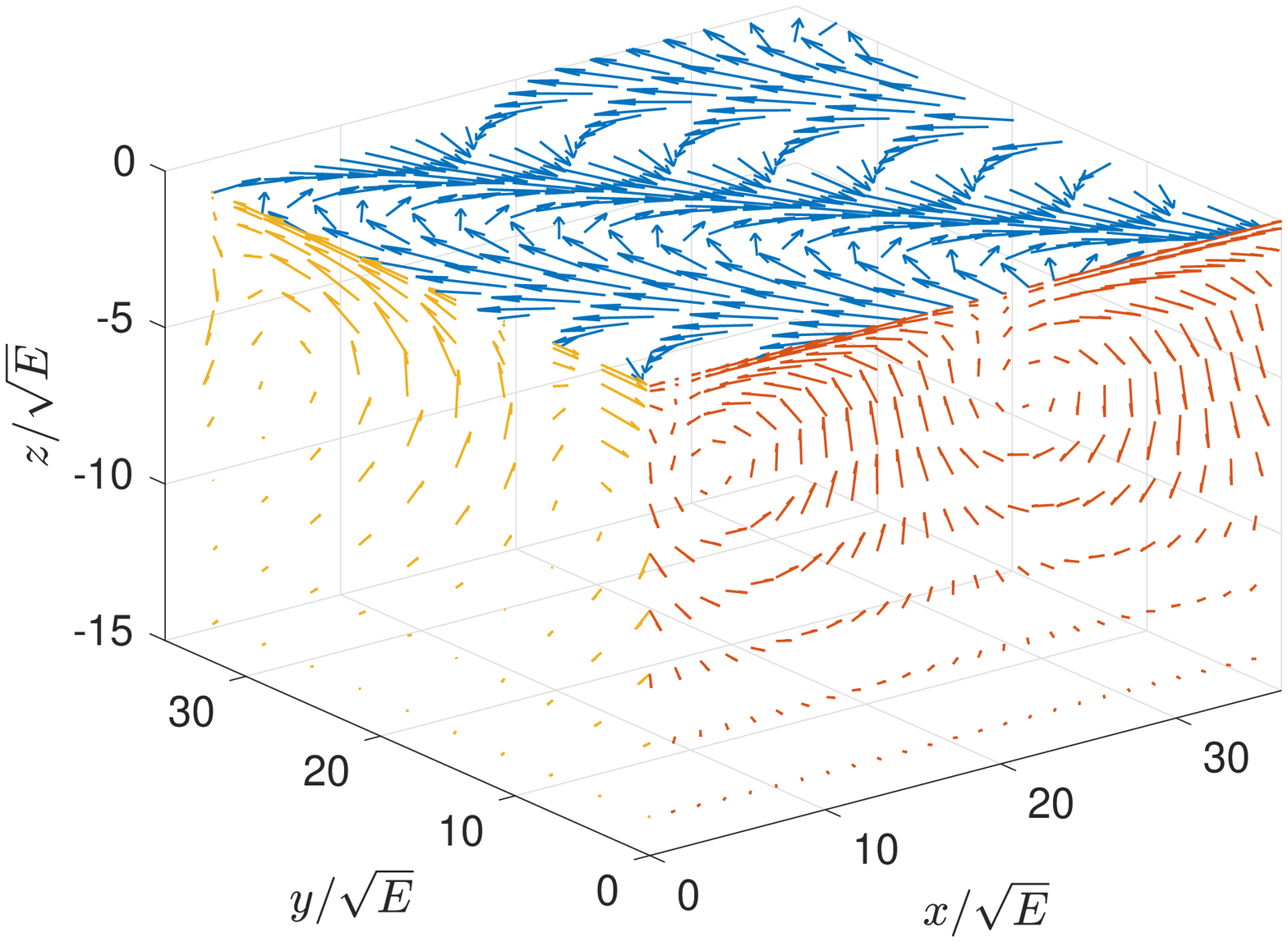}
\end{center}
\caption{Marginally stable eigenmode at the instability threshold for $\text{E}=10^{-8}$ (left) and $\text{E}=10^8$ (right). We represent the projection of the velocity vector onto each side of the cube.\label{fig:snapshots}}
\end{figure}

We consider the stability of the Ekman-Stokes spiral to infinitesimal perturbations. The base-flow being invariant along $x$ and $y$, we focus on a single Fourier mode of perturbation in these directions, together with exponential growth/decay in time. The velocity field reads:
\begin{align}
{\bf u} \left( x, y, z, t \right) = \begin{pmatrix}
u(z) \\
v(z) \\
0
\end{pmatrix} + \begin{pmatrix}
\tilde{u}(z) \\
\tilde{v}(z) \\
\tilde{w}(z)
\end{pmatrix} e^{i (k_x x + k_y y + \gamma t)} + c.c. \, ,
\end{align}
where the $\tilde{\cdot}$ denote infinitesimal perturbations, $k_x$ and $k_y$ are the wavenumbers of the perturbation along $x$ and $y$, and $c.c.$ is the complex conjugate of the second term. The angular frequency $\gamma$ is complex, and if its imaginary part $\text{Im}\left(\gamma\right)$ is negative then the perturbation grows exponentially with time. Upon linearizing the Craik-Leibovich equation (\ref{eq:rotatingCLadim}) around the base-flow (\ref{eq:base_flow_u}-\ref{eq:base_flow_v}), we obtain two coupled equations for $\tilde{w}$ and $\tilde{\omega}_z = i k_x \tilde{v} - i k_y \tilde{u}$:
\begin{align}
i\gamma \Delta \tilde{w} = & - \Big[ i k_x \left( u + u_s \right) + i k_y v \Big] \Delta \tilde{w} + \tilde{w} \partial_{zz} \Big[ i k_x u + i k_y v \Big] \nonumber \\
& + i k_y \tilde{\omega}_z \partial_z u_s - \partial_z \tilde{\omega}_z + \text{E} \, \Delta^2 \tilde{w} \, , \label{eq:linw} \\
i\gamma \, \tilde{\omega}_z = & - \Big[ i k_x \left( u + u_s \right) + i k_y v \Big] \tilde{\omega}_z - \tilde{w} \partial_z \Big[ i k_x v - ik_y u \Big] \nonumber \\
& + \partial_z \tilde{w} + \text{E} \, \Delta \tilde{\omega}_z \label{eq:linomz}\, . 
\end{align}
The boundary conditions for the perturbations are:
\begin{align}
\tilde{w}  = \partial_z \tilde{u} = \partial_z \tilde{v} = \partial_z \tilde{\omega}_z  =  0 \qquad \text{ at } z=0 \text{ and } z \to -\infty \, . \label{eq:BClin}
\end{align}
Equations (\ref{eq:linw}-\ref{eq:linomz}) with the boundary conditions (\ref{eq:BClin}) represent an eigenvalue problem for the complex eigenvalue $\gamma$, that we solve numerically using a Chebyschev collocation method. In practice, we replace the boundary conditions at infinite depth by a stress-free boundary at some depth $z=-H$, choosing a large enough $H$ for the results to be independent of $H$. We focus on the eigenvalue with the lowest imaginary part, after minimization over the horizontal wavenumbers $k_x$ and $k_y$, as it corresponds to the least stable (or most unstable) eigenmode. For a given value of the Ekman number, the imaginary part of this eigenvalue becomes negative above a threshold value $\text{Ro}_c$ of the Rossby number, which indicates a linear instability. In figure \ref{fig:linstab}a, we plot this critical Rossby number as a function of the Ekman number. $\text{Ro}_c$ is always of the order of unity, and it tends to limiting asymptotic values in both limits $\text{E} \to \infty$ and $\text{E} \to 0$. The numerical eigenvalue problem becomes particularly stiff in the latter limit, which calls for a theoretical confirmation. The following section is thus devoted to the analytical determination of the $\text{E}\to \infty$ and $\text{E}\to 0$ asymptotic values of the threshold Rossby number. In Fig. \ref{fig:linstab}b, we show the wavevector and angular frequency $\text{Re}(\gamma)$ of the most unstable mode at the threshold of the instability ($\text{Ro} = \text{Ro}_c, \text{Im}\left(\gamma\right) = 0$). The angular frequency is always nonzero, indicating a Hopf bifurcation, with a value close to $f$. The wavevector is predominantly along $x$ for low Ekman number, which corresponds to cellular motion, the axes of the rolls being horizontal and perpendicular to the direction of wave propagation. By contrast, for large Ekman number $k_x$ and $k_y$ are comparable, which corresponds to rolls whose axes are horizontal and form an angle close to 45$^\circ$ with the direction of wave propagation. In figure \ref{fig:snapshots}, we show the detailed structure of the eigenmode in both limits.


\section{Asymptotic behaviour for low and large Ekman number \label{sec:asympt}}

To gain further understanding of the instability mechanism, we focus on the asymptotic limits of low and large Ekman number. We will see that, at large Ekman number, the problem reduces to the stability analysis of a standard Ekman spiral driven by the effective surface stress (\ref{eq:stress}), and a connection can be made with the literature on Ekman spiral instabilities. By contrast, in the low-$\text{E}$ limit one needs to retain both the body force ${\bf u}_s \times 2 {\bm \Omega}$ and the effective surface-stress (\ref{eq:stress}), and a novel instability arises. 

\subsection{Large Ekman number limit, $\text{E} \to \infty$.}

In the large-Ekman-number limit, the base-flow (\ref{eq:base_flow_u}-\ref{eq:base_flow_v}) reduces to a standard Ekman spiral driven by the wave-induced surface stress (\ref{eq:stress}). The norm of the velocity vector at the free surface is $V = 4 \pi U_s \sqrt{\text{E}}$, in dimensional form. Through a systematic expansion in powers of $\text{E}^{-1/2}$, we show in appendix \ref{apphighEk} that the Stokes-drift profile disappears from the linearized equations, which reduce to the standard linear stability problem of the Ekman spiral. These equations depend on the parameter $\text{Ro}$ only, allowing us to compute numerically the asymptotic threshold value $\text{Ro}_c$ in the limit of large Ekman number. We solved this linear stability problem numerically and obtained the asymptotic threshold value $\text{Ro}_c|_{\text{E}\to \infty}=0.67$.

We can compare this value to the instability threshold computed by Faller \& Kaylor \cite{Faller} for the standard Ekman spiral. They obtain an instability for $Re \geq 12$, where $Re$ is a Reynolds number based on the norm $V$ of the base-flow velocity at the surface and the Ekman-depth $\sqrt{\nu/f}$: $Re=V/\sqrt{\nu f/2}=4\pi \sqrt{2} \,\text{Ro}$. Their instability threshold thus corresponds to $\text{Ro}_c=0.67$ using our notations, which fully confirms the value that we obtained numerically (and reported in figure \ref{fig:linstab}a). This validates our numerical computations, and the fact that the effective body-force $ {\bf u}_s \times 2 {\bm \Omega}$ is negligible in this limit, the flow being predominantly an Ekman spiral driven by the wave-induced surface stress.

\subsection{Weak viscosity regime, $\text{E} \to 0$. \label{sec:lowEk}}

In the low-Ekman-number limit, the base-flow evolves over two-length scales: the ${\cal O}(1)$ characteristic depth of the Stokes drift, and the ${\cal O}(\text{E}^{1/2})$ Ekman-layer thickness. Accordingly, the eigenmode can be computed in the asymptotic limit $\text{E}\to 0$ through a boundary layer expansion. It consists of an outer solution at the scale ${\cal O}(1)$, plus some small boundary layer corrections on a scale ${\cal O}(\text{E}^{1/2})$ near $z=0$. We introduce a small parameter $\epsilon=\sqrt{\text{E}}$ and the following ansatz for the form of the eigenmode:
\begin{eqnarray}
\tilde{w} & = &  \bar{w}_0(z)+\epsilon^2 \, [\bar{w}_2(z)+\hat{w}_2(z/\epsilon)] + {\cal O}(\epsilon^3) \label{ansatz1}\\
\tilde{\omega}_z & = &  \bar{\omega}_0(z) + \epsilon \, \hat{\omega}_1(z/\epsilon) + \epsilon^2 \, [\bar{\omega}_2(z)+\hat{\omega}_2(z/\epsilon)] + {\cal O}(\epsilon^3) \, . \label{ansatz2}
\end{eqnarray}
The quantities with a $\bar{\cdot}$ denote the outer solution, while the quantities with a $\hat{\cdot}$ denote boundary layer corrections. Both the fields with a $\bar{\cdot}$ and a $\hat{\cdot}$, and their derivatives, are ${\cal O}(1)$ when their argument is ${\cal O}(1)$. All the boundary layer corrections (the fields with a $\hat{\cdot}$) vanish when their argument is large, i.e., when $z/\epsilon \to -\infty$.

\subsubsection{Outer solution}
The outer region corresponds to $z={\cal O}(1)$, $z/\epsilon \to -\infty$. In this limit the boundary layer corrections disappear and the base flow reduces to:
\begin{align}
u = & - \text{Ro} \, e^{4 \pi z} + O(\epsilon^4) \, , \label{eq:baseflowuEk0}\\
v = &  \, \epsilon^2 16 \pi^2 \text{Ro} \, e^{4 \pi z} + O(\epsilon^5) \, . \label{eq:baseflowvEk0}
\end{align}
It consists of an anti-Stokes flow (\ref{eq:baseflowuEk0}) and a much weaker cross-flow (\ref{eq:baseflowvEk0}). At order ${\cal O}(\epsilon^0)$, equations (\ref{eq:linw}-\ref{eq:linomz}) yield:

\begin{eqnarray}
i \gamma \left( \frac{d^2}{dz^2} - k_x^2 \right) \bar{w}_0(z) + 16 \pi^2 i k_x \text{Ro}\, e^{4 \pi z} \bar{w}_0(z)  +  \bar{\omega}_0'(z) & = & 0 \, ,  \label{O0w}\\
i \gamma \bar{\omega}_0(z) - \bar{w}_0'(z) & = &  0 \, ,
\end{eqnarray} 
where the symbol $'$ denotes a derivative with respect to $z$. Keeping only the solution that decays for $z\to -\infty$, we obtain:
\begin{eqnarray}
\bar{w}_0 \left( z \right) & = & A \, \mathcal{J}_{a} \left[ b  \,e^{2 \pi z} \right] \Gamma\left( 1 + a \right) \label{soln_u0} \\
\bar{\omega}_0 \left( z \right) & = & A \, \frac{\pi \, b \, e^{2\pi z}}{ i \gamma} \Big( \mathcal{J}_{a - 1} \left[ b \, e^{2 \pi z} \right] - \mathcal{J}_{a + 1} \left[ b \, e^{2 \pi z} \right]  \Big) \Gamma\left( 1 + a \right) \, , \label{soln_o0_pathi}
\end{eqnarray} 
where $a = k_x \gamma/( 2 \pi \sqrt{\gamma^2 - 1})$, $b = 2 \sqrt{\gamma k_x Ro}/\sqrt{\gamma^2 - 1}$, $\mathcal{J}$ denotes a Bessel function of the first kind, $\Gamma$ denotes the standard Gamma function, and $A$ is the arbitrary complex amplitude of the eigenmode.
At order ${\cal O}(\epsilon^2)$, equations (\ref{eq:linw}-\ref{eq:linomz}) read:
\begin{eqnarray}
 & & i \gamma \left( \frac{d^2}{dz^2} - k_x^2 \right) \bar{w}_2(z)+ 16 \pi^2 i k_x \text{Ro}e^{4 \pi z} \bar{w}_2(z)  +  \bar{\omega}_2'(z)  =  \left( \frac{d^2}{dz^2} - k_x^2 \right)^2 \bar{w_0},  \\
& &  i \gamma \bar{\omega}_2 - \bar{w}_2'(z)  =   - 64 \pi^3 i k_x \text{Ro}e^{4 \pi z} \bar{w}_0 + \left( \frac{d^2}{dz^2} - k_x^2 \right)\bar{\omega}_0(z) \, .
\end{eqnarray} 
The solution to the homogeneous equation would just correspond to a slight modification of the amplitude $A$ in the ${\cal O}(1)$ outer solution, so we can discard it (or include it into $A$). In the following we only need the $\bar{w}_2$ response to the forcing terms on the right-hand side of these two equations, evaluated at $z=0$. We obtain:
\begin{eqnarray}
\nonumber \bar{w}_2(0)& = & \frac{4 \, i \, \pi^2 A}{2 a \gamma}  \mathcal{J}_{-a} \left[ b \right] \Gamma \left( 1- a \right)  \left\{ \Gamma^2 \left( 1+ a \right) \frac{b^2}{6} \left[ 3(a^2 + 2)+p (b^2- 2 -4 a^2+3 \frac{k_x^2}{4 \pi^2} ) \right] \mathcal{J}^2_{a - 1} \left[ b \right]  \right. \\ 
 & & \left. +\frac{a}{2}(p-1)(a^2 - \frac{k_x^2}{4 \, \pi^2} )  \left( \frac{b}{2}\right)^{2\, a} {_2F_3} \left[ \left(\frac{1}{2} + a, a \right),  \left(1 + a, 1 + a ,1 +  2 \, a \right) ,  -b^2 \right]      \right\} \, , \label{w20}
\end{eqnarray} 
where $p = (\gamma^2 + 1)/(\gamma^2 - 1)$, and ${_2F_3}$ denotes a generalized hypergeometric function.

\subsubsection{Inner expansion and boundary conditions}

Before considering the inner expansion, let us focus on the boundary condition $\tilde{w}(0)=0$, which at order ${\cal O}(1)$ simply yields $\bar{w}_0(0)=0$. This leads to a first relation between the eigenvalue $\gamma$, $k_x$ and $Ro$ in terms of $a, b$ as:
\begin{eqnarray}
\mathcal{J}_{a} \left[ b \right]  = 0. \label{constraint1}
\end{eqnarray}
One can check that when this constraint is satisfied we also have $\bar{\omega}_0'(0)=0$, which we substitute into (\ref{O0w}) to obtain $\bar{w}_0''(0)=0$. The constraint (\ref{constraint1}) is not sufficient to fully determine the eigenvalue $\gamma$. One needs to consider the inner expansion to get an additional constraint. We thus introduce the inner variable $Z=z/\epsilon$ before expanding the ansatz (\ref{ansatz1}-\ref{ansatz2}) in the inner region, $z \ll 1$ and $Z={\cal O}(1)$. In this region the outer solutions are expanded as, e.g., $\bar{w}_0(z)= \bar{w}_0(0) + z \bar{w}_0'(0)+ z^2\bar{w}_0''(0)/2 + \mathcal{O}(\epsilon^3)= \bar{w}_0(0) + \epsilon Z \bar{w}_0'(0) +\epsilon^2 Z^2 \bar{w}_0''(0)/2 + \mathcal{O}(\epsilon^3)$, and the fields read:
\begin{eqnarray}
\tilde{w} & = &\epsilon \, Z \bar{w}'_0(0)+\epsilon^2 [\bar{w}_2(0)+\hat{w}_2(Z)] + \mathcal{O}(\epsilon^3) \, ,\\
\tilde{\omega}_z & = & \bar{\omega}_0(0) +\epsilon \, \hat{\omega}_1(Z) + \mathcal{O}(\epsilon^2) \, ,
\end{eqnarray}
where we have simplified the expressions using $\bar{w}_0(0)=0$, $\bar{\omega}_0'(0)=0$ and $\bar{w}_0''(0)=0$. Differentiating these expressions with respect to $z$, remembering that $\frac{d}{dz}=\frac{1}{\epsilon} \frac{d}{dZ}$, the boundary conditions (\ref{eq:BClin}) evaluated order-by-order yield:
\begin{eqnarray}
\bar{w}_2(0) + \hat{w}_2(0) =0 , \qquad \hat{w}_2''(0)=0 , \qquad \text{and} \qquad \hat{\omega}_1'(0)=0 \, . \label{BCinner}
\end{eqnarray}
The combinations of the Stokes drift, the base flow and its derivatives appearing in the linearized equations take the following form in the inner region:
\begin{eqnarray}
u(z)+u_s & = & \epsilon \, 8 \pi \text{Ro}\, e^{\frac{Z}{\sqrt{2}}} \cos\left(\frac{Z}{\sqrt{2}}-\frac{\pi}{4} \right) + {\cal O}(\epsilon^3) \, , \\
\partial_z v & = & 8 \pi \text{Ro} \, e^{\frac{Z}{\sqrt{2}}}  \sin\left( \frac{Z}{\sqrt{2}} \right)  + {\cal O}(\epsilon^2) \, .
\end{eqnarray}
The governing equations for $\hat{w}_2$ and $\hat{\omega}_1$ are obtained by considering equation (\ref{eq:linw}) at order ${\cal O}(1)$ and equation (\ref{eq:linomz}) at order ${\cal O}(\epsilon)$, in the inner region:
\begin{eqnarray}
\hat{w}_2''''(Z)-i\gamma \hat{w}_2''(Z) - \hat{\omega}_1'(Z) & = & 0 \, , \label{eq:w2}\\
 - \hat{\omega}_1''(Z) + i \gamma \hat{\omega}_1(Z) -   \hat{w}_2'(Z) & = & -\bar{\omega}_0(0) \, 8 \pi i k_x \text{Ro}\, e^{\frac{Z}{\sqrt{2}}} \label{eq:om1}\\
\nonumber &  \times & \left[ \cos\left(\frac{Z}{\sqrt{2}}-\frac{\pi}{4} \right)  +  i \gamma Z \sin\left(\frac{Z}{\sqrt{2}} \right)  \right] \, .
\end{eqnarray}
We integrate the first equation from $-\infty$ to $Z$, to remove one order of differentiation from each term. This yields an expression for $\hat{\omega}_1(Z)$ in terms of $\hat{w}_2'''(Z)$ and $\hat{w}_2'(Z)$, which we insert into the second equation to obtain an ODE for $\hat{w}_2(Z)$ only. Integrating the resulting equation from $-\infty$ to $Z$ allows us to remove again one order of differentiation, and we finally obtain:
\begin{eqnarray}
\nonumber & & \hat{w}_2''''(Z)-2i\gamma \hat{w}_2''(Z)+(1-\gamma^2)\hat{w}_2(Z) = \\
& & \qquad 4 \pi k_x \text{Ro}\bar{\omega}_0(0) \, e^{\frac{Z}{\sqrt{2}}} \left[ (-2+\sqrt{2}Z) \gamma \cos\left( \frac{Z}{\sqrt{2}}\right) + (2i-\sqrt{2}\gamma Z)  \sin \left( \frac{Z}{\sqrt{2}}\right) \right] \, . \label{eqtotw2}
\end{eqnarray}
Inserting the boundary conditions (\ref{BCinner}) into equation (\ref{eq:w2}) evaluated at $Z=0$ yields an additional boundary condition on $\hat{w}_2$: $\hat{w}_2''''(0)=0$. Evaluating equation (\ref{eqtotw2}) at $Z=0$, remembering that $\hat{w}_2''''(0)=0$, $\hat{w}_2''(0)=0$ and $\hat{w}_2(0)=-\bar{w}_2(0)$, finally gives:
\begin{eqnarray}
(\gamma^2-1)\bar{w}_2(0) = -8\pi \gamma k_x \text{Ro} \, \bar{\omega}_0(0) \, . \label{addBC}
\end{eqnarray}
It is worth stressing the fact that we have obtained an additional constraint on the outer solution, (\ref{addBC}), without even solving explicitly for the inner correction. The expression of $\bar{w}_2(0)$ is given in (\ref{w20}), while $\bar{\omega}_0(0)$ is obtained by evaluating (\ref{soln_o0_pathi}) at $z=0$. Substitution into (\ref{addBC}) finally leads to the second relation between $\gamma$, $k_x$ and $Ro$:
\begin{eqnarray}
\nonumber & & \mathcal{J}_{-a} \left[ b \right] \mathcal{J}_{a - 1} \left[ b \right]  \frac{2 \, \pi^2 \, b^2}{3 a k_x^2} \left[ 2 a^2 \left( - 2 +  b^2 - 4 a^2 \right) + \frac{k_x^2}{4 \pi^2} \left( 8 + 13 a^2 - b^2 \right) - \frac{3 \, k_x^4}{16 \pi^4} \right] \\
\nonumber & &  + \frac{4 \pi^2}{k_x^2} \left( \frac{b}{2} \right)^{2 a} \frac{ \left( - a^2 +\frac{k_x^2}{4 \pi^2} \right)^2 }{\Gamma[1 + a]^2 } \frac{\mathcal{J}_{- a} \left[ b \right]}{\mathcal{J}_{a - 1} \left[ b \right]} \; \; {}_2F_3\Big[ \frac{1}{2} + a, a; 1 + a, 1 + a, 1 + 2 a; \Bigg. \Bigg.   - b^2 \Big]\\
 & & = \frac{2 b^3 \sin \left( \pi a \right) }{\pi a} \, .  \label{constraint2}
\end{eqnarray}

\subsubsection{Critical Rossby number and most unstable mode}

\begin{figure}
\includegraphics[width=8cm]{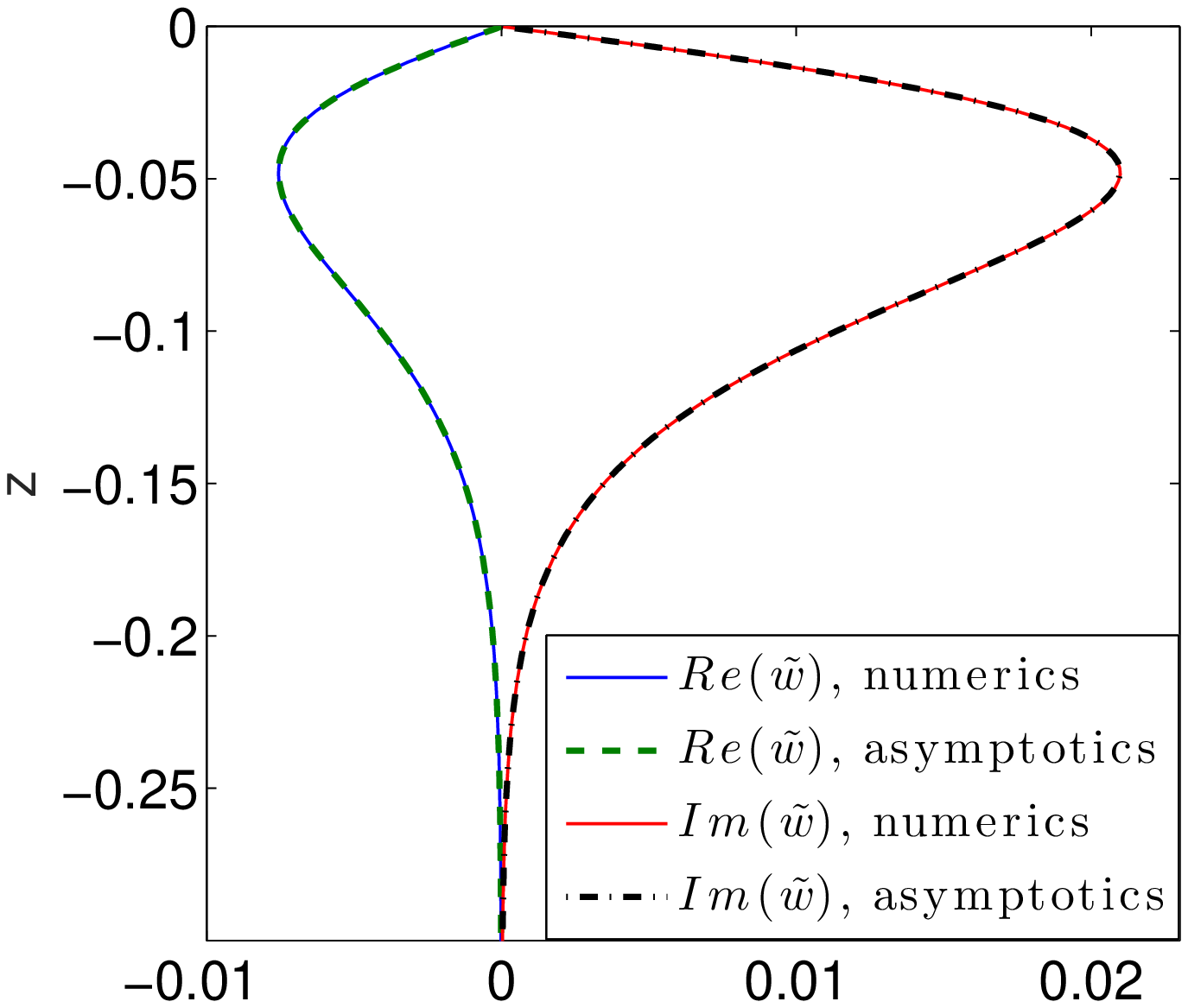}
\includegraphics[width=8cm]{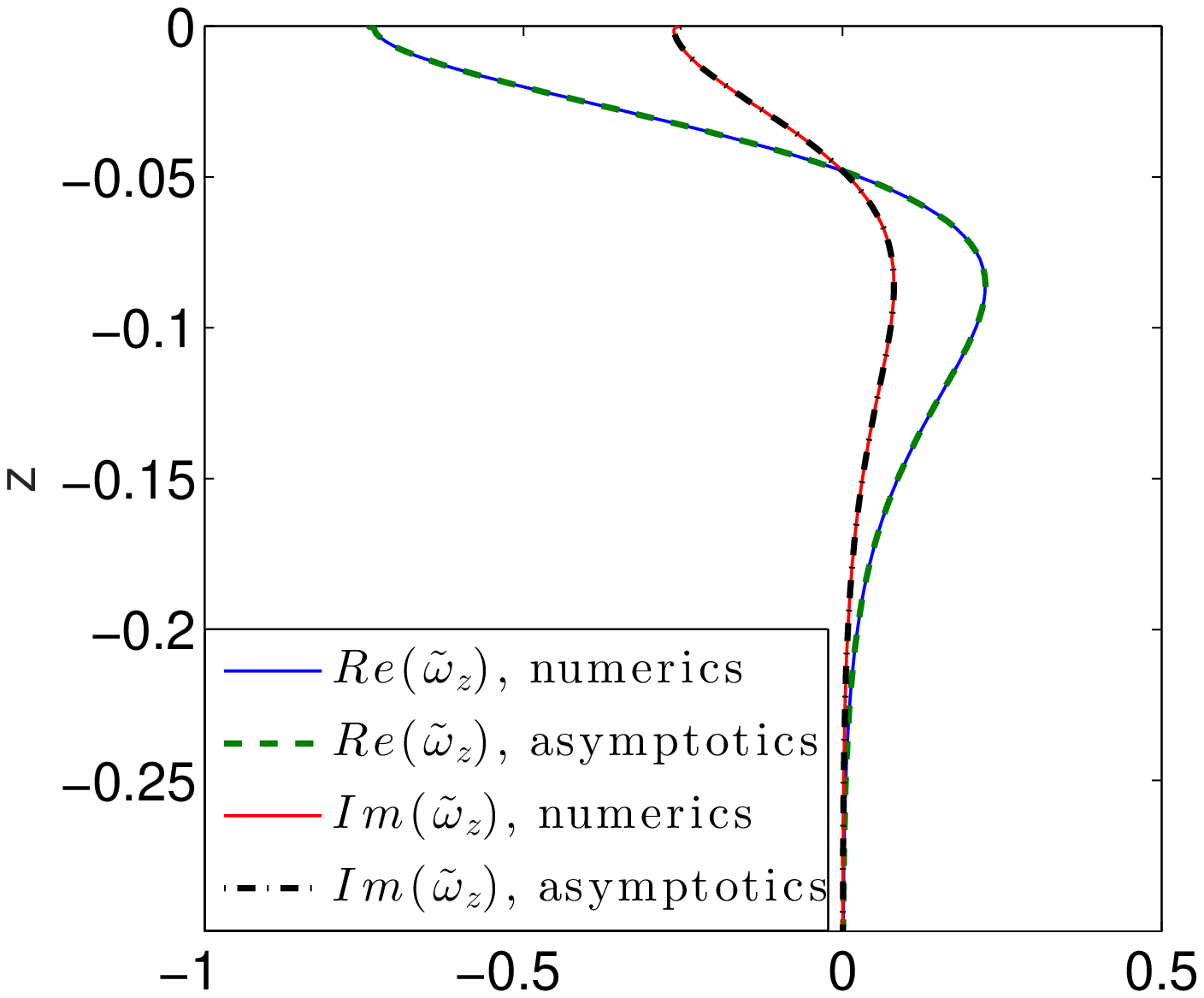}
\caption{Eigenmode in the low-$\text{E}$ regime (left: $\tilde{w}$, right: $\tilde{\omega}_z$). The analytical expression in the asymptotic limit $\text{E} \to 0$ (dashed lines) accurately matches the eigenmode computed numerically for $\text{E} = 10^{-8}$, $\text{Ro} = 0.24$ and $k_x = 12$ (solid lines).}
\label{fig:uoprofile}
\end{figure}

The two implicit relations (\ref{constraint1}) and (\ref{constraint2}) allow for the determination of the eigenvalue $\gamma$. The critical Rossby number is obtained by demanding that $\gamma$ be purely real and minimizing over $k_x$.
The theoretical minimum is found at $k_x = 12.67$, with an associated critical Rossby number $\text{Ro}_c = 0.23\pm 0.01$, the eigenvalue -- the angular frequency of the Hopf bifurcation -- being $\gamma = 1.059$. These asymptotic values are represented as dashed lines in the low-$\text{E}$ region of Fig.~\ref{fig:linstab} and fully confirm the numerical results. In Fig.~\ref{fig:uoprofile}, the asymptotic eigenmode is compared to the one computed numerically for $\text{E} = 10^{-8}$, $\text{Ro} = 0.24$ and $k_x = 12$. The agreement is excellent, the numerical results being accurately captured by the asymptotic analysis at low Ekman number.


\section{Discussion\label{sec:discussion}}

We have shown that the Ekman-Stokes spiral driven by surface waves propagating in a rotating frame is unstable when the Stokes drift of the waves is fast enough. For large Ekman number, the stability problem reduces to that of a standard Ekman spiral driven by the wave-induced surface stress (instead of a wind stress in the classical case). By contrast, for low Ekman number the Stokes drift profile directly affects both the instability threshold and structure of the eigenmode, which we captured through a boundary-layer expansion valid in the asymptotic limit $\text{E} \to 0$. We stress the fact that, in that limit, the instability differs strongly from the inviscid instability reported by Craik \cite{Craik1977}: first, Craik's instability induces a cellular flow structured in the spanwise direction, whereas the present instability induces preferentially a streamwise-varying cellular flow. Second and more importantly, the low-$\text{E}$ bulk anti-Stokes flow corresponds to ``Case II'' described in section 3 of Ref.~\cite{Craik1977}; because the shear in the Stokes drift profile and in the anti-Stokes Eulerian flow have opposite signs, the flow is stable with respect to Craik's instability mechanism.


\subsection{Expressing the instability threshold as a critical Lagrangian Reynolds number}

Although we have expressed the instability threshold in terms of a critical Rossby number throughout this study, it is worth stressing the fact that, in both the large- and low-$\text{E}$ asymptotic limits, this threshold can be recast as a critical Reynolds number associated to the Lagrangian flow ${\bf u}_L={\bf u}+{\bf u}_s$. Indeed, for large Ekman number, the base flow reduces to a standard Ekman spiral. The Eulerian flow is much faster than the Stokes drift, so that the latter can be neglected in the expression of the Lagrangian velocity ${\bf u}_L$. The Lagrangian flow thus has a typical dimensional surface speed $U_L \sim \sqrt{\text{E}} U_s$ and a dimensional length-scale $L_L \sim \sqrt{\nu/f}=\sqrt{\text{E}} \lambda$. Using $\nu = \text{E} \lambda^2 f$, the Reynolds number associated to this Lagrangian base flow is $Re^{(L)}=U_L L_L / \nu = Ro$. Similarly, for low Ekman number the base flow reduces to (\ref{eq:baseflowuEk0}-\ref{eq:baseflowvEk0}). The corresponding Lagrangian flow is ${\bf u}_L={\bf u}+{\bf u}_s \simeq \text{E} \, 16 \pi^2 \text{Ro}e^{4 \pi z} {\bf e}_y$. It varies on a typical dimensional length $L_L \sim \lambda$, with a dimensional speed of order $U_L \sim \text{E} \, \text{Ro}\, \lambda f$. The dimensional viscosity being $\nu = \text{E} \, \lambda^2 f$, the Reynolds number associated to the Lagrangian flow is, again, $Re^{(L)}=U_L L_L/ \nu = \text{Ro}$. In both the low- and large-$\text{E}$ limits, we conclude that the Ekman-Stokes spiral becomes unstable when the Reynolds number associated to the Lagrangian velocity profile exceeds a threshold value independent of $E$.

 \subsection{Relevance in an oceanographic context}

The combination of surface waves and global rotation arises when swell propagates at the ocean surface. Of course, the pilot problem considered throughout this study lacks several important ingredients of oceanic dynamics, such as density stratification, a horizontal component of the global rotation vector, or spatio-temporal fluctuations of the swell. Nevertheless, it indicates rather clearly that the anti-Stokes flows referred to in the oceanographic literature are unstable in standard oceanic conditions. Indeed, with a typical swell wavelength of $100$m, the Ekman number is approximately $\text{E}  \, {\simeq} 10^{-6}$. This is well into the low-Ekman-number regime where the base flow is dominated by the anti-Stokes flow, $-{\bf u}_s$, plus a small Ekman-spiral correction.  
With a typical wave amplitude $a_w \simeq 1$m, the Stokes drift amplitude is $U_s \simeq 5.\, 10^{-2}$m.s$^{-1}$ and the Rossby number is $\text{Ro}\simeq 5$, well above the threshold value $\text{Ro}_c=0.23$ for instability. At such distance from threshold, we found several unstable modes numerically, with growth rates as large as $700\times f \simeq 0.1\text{s}^{-1}$. Preliminary nonlinear simulations at large distance from threshold indicate that the flow can become time-dependent and behave chaotically.

The instability presented here is thus the first of a series of instabilities leading to chaotic and possibly turbulent flows. However, we argue that this mechanism cannot extract much energy from the wave field. Indeed, we will now show that the power dissipated by the mean flow -- and extracted from the wave field -- remains negligible as compared to the power dissipated through standard viscous damping of the propagating surface gravity waves. Focusing on dimensional quantities in this discussion section, we form the Lagrangian kinetic energy budget by taking the dot product of equation (\ref{eq:CL_rotatingframe}) with ${\bf u}+{\bf u}_s$:
\begin{equation}
\partial_t \left[{\bf u}\cdot \left(\frac{{\bf u}}{2}+{\bf u}_s\right) \right] = -{\bm \nabla}\cdot [p ({\bf u}+{\bf u}_s)] + \nu ({\bf u}+{\bf u}_s)\cdot  \Delta {\bf u} \, . \label{eq:local_budget}
\end{equation}
Consider this equation in a Cartesian domain ${\cal D} = [0,L_x]\times[0,L_y]\times[-H,0]$, with $kH \gg 1$, periodic boundary conditions in $x$ and $y$, and an impenetrable stress-free boundary at $z=-H$. Integrate (\ref{eq:local_budget}) over ${\cal D}$, perform a few integrations by parts using the boundary conditions (\ref{eq:stress}), multiply by the density $\rho$ and divide by $L_x L_y$ before time-averaging, to get:
\begin{equation}
P_{\text{diss}} = \frac{\rho \nu}{L_x L_y} \int_{\cal D} \left< |{\bm \nabla} {\bf u}|^2\right> \mathrm{d}{\bf x} = 2k \rho \nu U_s^2 + \frac{4k^2 U_s \rho \nu}{L_x L_y}  \int_{\cal D} \left< u \right> e^{2kz} \mathrm{d}{\bf x} \, ,
\end{equation}
where $P_{\text{diss}}$ denotes the power dissipated by the mean flow ${\bf u}$ per unit area of fluid surface (in kg.s$^{-3}$), and $\left< \cdot \right>$ is time average. We bound the second term on the right-hand side by extracting from the integral the spatial maximum of $\left< u \right>$:
\begin{equation}
P_{\text{diss}} \leq 2 k \rho \nu U_s \left[ U_s + \text{max}_{{\bf x} \in \cal D} \left< u \right> (1-e^{-2kH})   \right] \, .
\end{equation}
After taking the limit $kH \to \infty$ and introducing the Froude number $\text{Fr}= \text{max}_{{\bf x} \in \cal D} \left< u \right> / c$, where $c$ is the phase speed of the waves, we obtain:
\begin{equation}
P_{\text{diss}} \leq 2 k \rho \nu U_s  c \left[ \frac{U_s}{c} + \text{Fr}  \right] = 2 P_{\text{waves}} [(a_w k)^2 + \text{Fr}]\, , \label{ineq}
\end{equation}
where $P_\text{waves}=k \rho \nu U_s  c$ denotes the standard viscous dissipation rate per unit horizontal area of the monochromatic wave field, and $a_w$ is the wave amplitude. In the Craik-Leibovich ordering, the mean flow arises at second order in wave slope, hence $\text{Fr} \sim (a_w k)^2 \ll 1$. The inequality (\ref{ineq}) thus shows that the dissipated power associated to the background mean flow is negligible as compared to the direct viscous dissipation of the surface-wave field, typically by a factor $(a_w k)^2$. We conclude that the transfer of energy from the wave field to the background mean flow is a negligible sink of wave energy in the present system, even in the fully nonlinear regime.

We close this study by proposing a scenario through which the present instability could have important consequences for near-surface particle dispersion in the ocean: assume that the succession of instabilities leads to a fully turbulent flow, where the kinetic energy dissipation rate per unit mass of fluid is of order $U^3/\lambda$, $U$ being the typical Eulerian velocity of the turbulent flow, which we assume to have a typical scale $\lambda$. Substituting this estimate into the left-hand side of (\ref{ineq}) yields $U/U_s \lesssim \left( {\nu}/{U_s \lambda} \right)^{1/3} \ll 1$ for typical ocean values, i.e., the Eulerian flow is much weaker than the anti-Stokes estimate $U \sim U_s$ of the initial base flow. The Lagrangian velocity is then of order ${\bf u}_L={\bf u}+ {\bf u}_s \simeq {\bf u}_s$, much larger than the anti-Stokes estimate ${\bf u}_L={\bf u}+ {\bf u}_s \simeq {\bm 0}$ associated with the initial base-flow. The instability thus triggers enhanced dispersion: particles and tracers are carried by the flow at a typical velocity $U_s$, instead of the vanishing Lagrangian velocity estimate associated with the anti-Stokes Eulerian flow. The central argument of this scenario is that the instability generates a genuinely turbulent flow that dissipates kinetic energy at a rate independent of viscosity, in line with the ``zeroth law of turbulence'' \cite{Frisch}. Whether this is the case can be investigated through direct numerical simulation of the fully non-linear equations, in the low-Ekman-number regime.

{\bf Acknowledgement}: We thank G. Chini, W. R. Young and S. Auma{\^i}tre for insightful discussions. This research is supported by the European Research Council under grant agreement 757239.

\bibliography{refs}{}
\bibliographystyle{plain}

\appendix

\section{Boundary conditions \label{appBC}}

We consider monochromatic surface gravity waves propagating towards $x>0$. We denote as $\eta(x,t)$ the surface displacement. Throughout this appendix, ${\bf u}=(u,v,w)$ denotes the full velocity field, including both the fast wave motion and the background mean flow, the latter arising at second order in wave slope. Angular brackets denote an average over one wave period only.
The kinematic boundary condition is:
\begin{align}
\partial_t \eta + \left( {\bf u} \cdot {\bm \nabla} \right) \eta = w |_\eta \, .
\end{align}
Averaging over one wave period, and expanding around $z = 0$ to second order in wave-slope, we obtain:
\begin{align}
\left < \partial_t \eta \right> + \left< u|_0 \partial_x \eta \right> \simeq \left<w |_0 \right> + \left< \eta \partial_z w|_0 \right> \, .
\end{align}
The first term on the left-hand side vanishes up to second-order in wave slope. The quadratic terms also vanish when one substitutes the various fields expressed for linear monochromatic waves, and we are simply left with $\left<w \right>|_0 = 0$, which is the standard no-penetration boundary condition for the background mean flow.

\begin{figure}
\includegraphics[width=14cm]{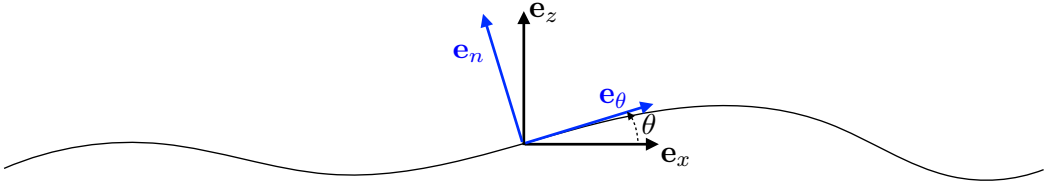}
\caption{Local frame attached to the interface: ${\bf e}_\theta$ denotes the vector tangent to the interface in the $xz$ plane. ${\bf e}_n$ denotes the vector perpendicular to ${\bf e}_\theta$ in the $xz$ plane, pointing upwards. $\theta$ denotes the local angle between the horizontal direction ${\bf e}_x$ and the tangent direction ${\bf e}_\theta$.\label{fig:schemappendix}}
\end{figure}

To address the consequences of a stress-free boundary condition, we denote as $\mu$ the dynamic viscosity and $\theta(x,t)$ the angle between the fluid surface and the horizontal, i.e., $\tan \theta = \partial_x \eta$. As sketched in figure {\ref{fig:schemappendix}}, we denote as ${\bf e}_\theta$ the local vector tangent to the interface in an $xz$ plane, pointing towards positive $x$. We denote as ${\bf e}_n$ the vector perpendicular to ${\bf e}_\theta$ in an $xz$ plane, pointing upwards. We introduce the stress tensor ${\bm \tau}$, such that $\tau_{ij}=-p \delta_{ij}+\mu(\partial_j u_i + \partial_i u_j)$, and ask for the local tangential stress at the (true) interface to vanish: ${\bf e}_n \cdot {\bm \tau}|_\eta  \cdot {\bf e}_\theta = 0$. Using the decompositions ${\bf e}_\theta= \cos \theta \, {\bf e}_x + \sin \theta \, {\bf e}_z$ and ${\bf e}_n= -\sin \theta \, {\bf e}_x + \cos \theta \, {\bf e}_z$, we obtain:
\begin{eqnarray}
0 & = & {\bf e}_n \cdot {\bm \tau}|_\eta  \cdot {\bf e}_\theta \\
\nonumber & = & - \sin \theta \, \cos \theta \, \tau_{xx}|_\eta + \cos^2 \theta \, \tau_{zx}|_\eta - \sin^2 \theta \, \tau_{xz}|_\eta + \sin \theta \, \cos \theta \, \tau_{zz}|_\eta \\
\nonumber & = & \frac{\sin{2 \theta}}{2} \, (\tau_{zz}|_\eta-\tau_{xx}|_\eta) + \cos{2\theta} \, \tau_{xz}|_\eta \, ,
\end{eqnarray}
where all the elements of $\tau$ are evaluated at the free surface $z=\eta$, and we have used $\tau_{xz}=\tau_{zx}$. Substituting the expressions of $\tau_{xx}|_\eta$ and $\tau_{zz}|_\eta$, we get:
\begin{eqnarray}
 0 & = & \mu \sin(2 \theta) (\partial_z w|_\eta - \partial_x u|_\eta) + \tau_{xz}|_\eta \cos(2 \theta) \, ,
\end{eqnarray}
where all the velocity derivatives are evaluated at the interface, $z=\eta$. Because we retain terms up to second order in wave slope only, the parenthesis $(\partial_z w - \partial_x u)$ above can be evaluated using the wave velocity field only, for which incompressibility implies $\partial_x u = -\partial_z w$. We finally obtain:
\begin{eqnarray}
\tau_{xz}|_\eta  = -2\mu \tan(2\theta) \, \partial_z w|_\eta \, .
\end{eqnarray}
In other words, $\tau_{xz}$ is nonzero because the surface is not flat and horizontal. Expanding the right-hand side to second-order in wave slope yields:
\begin{eqnarray}
\tau_{xz}|_\eta \simeq - 4 \mu \theta \partial_z w|_0 \simeq -4 \mu \partial_x(\eta) \partial_z w|_0 \, . \label{appeq2}
\end{eqnarray}
Averaging (\ref{appeq2}) over one period and substituting the expression of $\partial_x\eta$ and $\partial_z w|_0 $ for small-amplitude monochromatic waves leads to:
\begin{eqnarray}
\left< \tau_{xz}|_\eta \right> \simeq 2 \mu k^2a_w^2 \sigma = \mu \partial_z({\bf u}_s)|_0 \cdot {\bf e}_x \, .
\end{eqnarray}
For small wave amplitude, the linearized equation of motion along $x$ is:
\begin{eqnarray}
\partial_t u = -g e^{k z} \partial_x \eta + \frac{1}{\rho} \partial_z \tau_{xz} \, .
\end{eqnarray}
Now perform a momentum budget by integrating this equation from $z=-2a_w$ (slightly below the surface) to $z=\eta$, before averaging over one wave period:
\begin{eqnarray}
\left< \int_{-2a_w}^\eta \partial_t u \mathrm{d}z \right>= -g  \left< \int_{-2a_w}^\eta  e^{k z} \partial_x \eta  \mathrm{d}z \right>+ \frac{1}{\rho} ( \left< \tau_{xz}|_\eta \right> -   \left< \tau_{xz}|_{-2a} \right> ) \, . \label{appmombud}
\end{eqnarray}
The left-hand-side term vanishes to second order in wave slope:
\begin{eqnarray}
\left< \int_{-2a_w}^\eta \partial_t u \mathrm{d}z \right>=\left< \partial_t \left( \int_{-2a_w}^\eta  u \mathrm{d}z \right) \right> - \left< \partial_t(\eta)  u|_\eta  \right> \simeq 0 -  \left< \partial_t(\eta)  u|_0  \right> \simeq - \left< w|_0  u|_0  \right> = 0\, ,
\end{eqnarray}
where the last equality is obtained by substituting the expressions of $u|_0$ and $w|_0$ for linear monochromatic waves. The first term on the right-hand side of equation \eqref{appmombud} also vanishes to second-order in wave slope:
\begin{eqnarray}
-g  \left< \int_{-2a_w}^\eta  e^{k z} \partial_x \eta  \mathrm{d}z \right> & = - \frac{g}{k} \left< (e^{k \eta} - e^{- 2 k a_w} ) \partial_x \eta \right> \simeq -g \left< (\eta + 2 a_w) \partial_x \eta \right>  \nonumber \\
& =   -g \partial_x \left< \frac{\eta^2}{2} \right> - 2 a_w g \partial_x \left<\eta \right> = 0 \, ,
\end{eqnarray}
so that (\ref{appmombud}) reduces to $ \left< \tau_{xz}|_\eta \right> =  \left< \tau_{xz}|_{-2a_w} \right> \simeq \left< \tau_{xz}|_{0} \right>$, to second-order in wave-slope. We finally obtain:
\begin{eqnarray}
\left< \tau_{xz}|_{0} \right> =  \mu \partial_z({\bf u}_s)|_0  \cdot {\bf e}_x \, ,
\end{eqnarray}
valid up to second order in wave slope, which yields the boundary condition (\ref{eq:stress}) for the background mean flow. This boundary condition also holds if one allows for a small decay of the waves, either in time or along the direction of wave propagation \cite{Dore,Weber}.

\section{Reduction to the standard Ekman spiral stability problem for $\text{E} \to \infty$ \label{apphighEk}}

In the large-Ekman-number limit, and throughout this appendix only, we introduce the small parameter $\epsilon=1/\sqrt{\text{E}}$ and the slow variable $Z=\epsilon z$ (not to be confused with the $\epsilon$ and $Z$ introduced in section \ref{sec:lowEk}). The base-flow is expanded in powers of $\epsilon$ as:
\begin{eqnarray}
u = & \epsilon^{-1} 4 \pi \text{Ro} \, e^{\frac{Z}{\sqrt{2}}} \cos \left( \frac{Z}{\sqrt{2}} - \frac{\pi}{4} \right) - \epsilon \frac{Ro}{4 \pi} e^{ \frac{Z}{\sqrt{2}}} \cos \left( \frac{Z}{\sqrt{2}} + \frac{\pi}{4} \right) + O(\epsilon^2) \, ,  \\
v = & \epsilon^{-1} 4 \pi \text{Ro}\, e^{\frac{Z}{\sqrt{2}}} \sin \left( \frac{Z}{\sqrt{2}} - \frac{\pi}{4} \right) - \epsilon \frac{Ro}{4 \pi} e^{\frac{Z}{\sqrt{2}}} \sin \left( \frac{Z}{\sqrt{2}} + \frac{\pi}{4} \right) + O(\epsilon^2) \, . 
\end{eqnarray}

The Stokes drift profile is negligible at this order of approximation and the base-flow reduces to the standard Ekman spiral, which is a function of $Z$ only. Correspondingly, the eigenfunctions will evolve on the Ekman-layer depth only. We scale the perturbations as $\tilde{w}=w_0(Z)$ and $\tilde{\omega}_z=\epsilon \, {\omega}_1(Z)$. Guided by the large-$\text{E}$ behaviour of $k_x$ and $k_y$ in figure \ref{fig:linstab}, we scale the horizontal wavenumbers as $k_x=\epsilon \bar{k}_x$ and $k_y=\epsilon \bar{k}_y$. The linearized equations (\ref{eq:linw}) and (\ref{eq:linomz}) then yield at leading-order in $\epsilon$:
\begin{eqnarray}
& & i \gamma \left( \partial_{ZZ} - \bar{k}_x^2 - \bar{k}_y^2 \right) w_0 + \partial_Z \omega_1 - \left( \partial_{ZZ} - \bar{k}_x^2 - \bar{k}_y^2 \right)^2 w_0 = - 4 \pi \text{Ro}\, e^{\frac{Z}{\sqrt{2}}} \Big[ i \bar{k}_x \cos \left( \frac{Z}{\sqrt{2}} - \frac{\pi}{4} \right)  \label{Ekinf:eqns_kadaisi_1} \\
& &  + i \bar{k}_y \sin \left(\frac{Z}{\sqrt{2}} - \frac{\pi}{4} \right) \Big] \left( \partial_{ZZ} - \bar{k}_x^2 - \bar{k}_y^2 \right) w_0 + 4 \pi \text{Ro}\, e^{\frac{Z}{\sqrt{2}}} \Big[ i \bar{k}_x \cos \Big( \frac{Z}{\sqrt{2}} + \frac{\pi}{4} \Big) + i \bar{k}_y \sin \left( \frac{Z}{\sqrt{2}} + \frac{\pi}{4} \right) \Big] w_0 \, ,\nonumber \\
& & i \gamma \omega_1 - \partial_Z w_0 - \left( \partial_{ZZ} - \bar{k}_x^2 - \bar{k}_y^2 \right) \omega_1 = - 4 \pi \text{Ro}\, e^{\frac{Z}{\sqrt{2}}} \Big[ i \bar{k}_x \cos \left( \frac{Z}{\sqrt{2}} - \frac{\pi}{4} \right) + i \bar{k}_y \sin \left( \frac{Z}{\sqrt{2}} - \frac{\pi}{4} \right) \Big] \omega_1  \label{Ekinf:eqns_kadaisi_2} \\
& &  - 4 \pi \text{Ro}\, e^{\frac{Z}{\sqrt{2}}} \Big[ i \bar{k}_x \sin \left( \frac{Z}{\sqrt{2}} \right) - i \bar{k}_y \cos \left( \frac{Z}{\sqrt{2}} \right) \Big] w_0 \, , \nonumber
\end{eqnarray}

with the boundary conditions $w_0(0) = 0$, $\partial_Z {\omega}_1 |_{Z=0} =0$, and $\partial_{ZZ} w_0 |_{Z=0} = 0$. We solved this eigenvalue problem numerically, looking for the critical Rossby number $\text{Ro}_c$ at which the imaginary part of $\gamma$ vanishes. After optimizing over $\bar{k}_x$ and $\bar{k}_y$, the threshold Rossby number obtained numerically is $\text{Ro}_c=0.664$. The angular frequency of the unstable mode at threshold is $\gamma=1.067$, while the horizontal wavenumbers are $k_x=0.175/\sqrt{\text{E}}$ and $k_y=0.143/\sqrt{\text{E}}$. The instability induces a travelling cellular flow, the axis of the cells making an angle close to $45^\circ$ with the direction of wave propagation. These results are fully compatible with those briefly mentioned in Faller \& Kaylor \cite{Faller}.



%

\end{document}